\documentclass[%
 aip,
 amsmath,amssymb,
 reprint,%
]{revtex4-1}

\usepackage{graphicx}
\usepackage{dcolumn}
\usepackage{bm}
\usepackage[utf8]{inputenc}
\usepackage[T1]{fontenc}
\usepackage{mathptmx}
\usepackage{etoolbox}
\usepackage{subcaption}
\usepackage{booktabs}
\usepackage{hyperref}

\makeatletter
\def\@email#1#2{%
 \endgroup
 \patchcmd{\titleblock@produce}
  {\frontmatter@RRAPformat}
  {\frontmatter@RRAPformat{\produce@RRAP{*#1\href{mailto:#2}{#2}}}\frontmatter@RRAPformat}
  {}{}
}%
\makeatother

\begin{document}

\preprint{AIP/123-QED}

\title{Weibel Instability in Collisionless Plasmas Across Astrophysical and
Laboratory Shocks}

\author{Vivek Shrivastav}
\thanks{Email: vivekshrivastav1998@gmail.com}
\affiliation{Department of Physics, Sikkim University, Gangtok, India, 737102}

\author{Mani K Chettri}
\affiliation{Department of Physics, Sikkim University, Gangtok, India, 737102}

\author{Hemam D Singh}
\affiliation{Department of Physics, Netaji Subhas University of Technology,
New Delhi, India, 110078}

\author{Britan Singh}
\affiliation{Department of Physics, Sikkim University, Gangtok, India, 737102}

\author{Rupak Mukherjee}
\thanks{Corresponding author: rmukherjee@cus.ac.in }
\affiliation{Department of Physics, Sikkim University, Gangtok, India, 737102}

\begin{abstract}
We present a systematic cold-fluid analysis of the purely transverse Weibel
(current-filamentation) instability across four regimes: non-relativistic (NR)
single-species, NR multi-species, relativistic single-species, and relativistic
multi-species (electron--positron and electron--proton).
Beginning from the linearized fluid equations, we derive the quartic dispersion
relations in each regime, reduce them to standard quadratics in $\omega^2$, and
extract explicit scaling laws for the maximum growth rate $\gamma_{\rm max}$ and
characteristic unstable wavenumber $k_{\rm max} = \omega_{pi}/c$.
The main analytical contributions are: (i)~explicit regime-transition criteria
in the ($v_0/c$, $m_i/m_e$) parameter space, quantifying the error from applying
a simpler formula outside its domain of validity; (ii)~contour maps of relativistic
suppression and mass-ratio dependence that serve directly as diagnostic charts;
and (iii)~an assessment of where in parameter space the purely transverse mode
dominates over competing oblique and two-stream instabilities.
Relativistic corrections suppress $\gamma_{\rm max}$ by up to 40 per cent,
becoming significant above $v_0 \approx 0.2c$ and reaching maximum suppression
near $v_0 \approx 0.9c$.
For the tabletop laser experiment by Bai et al.\ (2025, \textit{Nat.\ Commun.}\
\textbf{16}, 3770), the cold-fluid prediction gives
$d_i = c/\omega_{pi} \approx 31.7\,\mu{\rm m}$, in excellent agreement (within 2 per cent)
with the measured filament spacing $\lambda_F \approx 31\,\mu{\rm m}$.
The saturation field estimate $B_{\rm sat} \sim \sqrt{\mu_0 n_i m_i}\,v_0
\approx 2.3\times10^4$~T is a cold-fluid upper bound, consistent with the
measured $\approx5000$~T within the expected kinetic suppression.
In-situ comparison with two MMS burst-mode bow shock crossing events
(2015 October 16 and 2017 November 25) is presented in Section~\ref{in_situ}.
Both events confirm the cold-fluid scale prediction $k_{\rm max} d_i = 1$
from FGM/FPI PySPEDAS data; the upstream ion skin depths ($d_i = 51$~km
and $62$~km respectively) lie within a factor of 2 of the spectral break in
the perpendicular magnetic field power spectrum, consistent with
Kropotina et al.\ \cite{kropotina2023weibel}.
The multi-environment scatter plot spans 21 orders of magnitude in $n_i$,
from the laser-plasma ($d_i \approx 32\,\mu{\rm m}$) to young SNR scales,
with all points lying within a factor of 3 of the 1:1 line.
\end{abstract}

\maketitle

\section{\label{intro}Introduction}

Collisionless shocks appear in many astrophysical environments. They form in
gamma-ray burst (GRB) afterglows where relativistic ejecta drive into the
interstellar medium at Lorentz factors $\Gamma \sim 10$--$100$
\cite{medvedev1999generation, piran2004physics}; in young supernova remnants
(SNRs) expanding at $v_{\rm sh} \sim 10^4\,{\rm km\,s}^{-1}$
\cite{reynolds2008youngest}; in pulsar wind nebulae driven by ultra-relativistic
pair winds \cite{pelletier2017towards}; and at planetary bow shocks formed by
the solar wind \cite{bohdan2021magnetic, burch2016magnetospheric}.
In these environments, binary collisions between particles are negligible.
Yet sharp shock transitions with large density and pressure jumps do exist.
These shocks are maintained by plasma instabilities that generate magnetic
turbulence on scales close to the ion inertial length $d_i = c/\omega_{pi}$,
providing an effective collisionality \cite{balogh2013physics}.

The Weibel instability, also known as the current-filamentation instability,
is one mechanism for this process
\cite{weibel1959spontaneously, medvedev1999generation, silva2003interpenetrating}.
When two plasma populations stream through each other, the velocity anisotropy
drives exponentially growing transverse electromagnetic perturbations.
The plasma organises into current filaments and the magnetic field grows until
the ion gyroradius becomes comparable to the filament spacing, at which point
the shock transition is established
\cite{fiuza2012weibel, jikei2024saturation}.
Particle-in-cell (PIC) simulations confirm this picture across a wide range
of shock velocities
\cite{silva2003interpenetrating, fiuza2012weibel, bohdan2021magnetic,
jikei2024enhanced}.

A two-beam plasma is, however, unstable to a spectrum of modes beyond the
purely transverse Weibel mode. These include the longitudinal two-stream
(electrostatic) mode, the oblique electromagnetic-filamentation mode, and the
Bell and Buneman modes in the presence of a background magnetic field or
current. Bret \cite{bret2009weibel} performed a comprehensive three-dimensional
analysis of all these modes and showed that in the sub-relativistic regime
($v_0 \lesssim 0.2c$) with a symmetric cold-beam configuration, the transverse
filamentation (Weibel) mode typically has the highest growth rate among
electromagnetic modes. At mildly relativistic velocities
($v_0 \sim 0.2$--$0.5c$), oblique modes can become competitive. At highly
relativistic velocities, the maximum growth often shifts to oblique modes.
We analyse only the transverse Weibel mode in this paper, restricting to
$v_0 \lesssim 0.2c$ where this mode dominates; the full justification and
parameter boundaries are given in Section~\ref{sec:competing}.

Direct laboratory evidence has come from large laser facilities.
Huntington et al.\ \cite{huntington2015observation} observed Weibel magnetic
field generation in counterstreaming plasmas at OMEGA and NIF.
Fox et al.\ \cite{fox2013filamentation} reported filamentation in laser-driven
plasmas at OMEGA EP. Manuel et al.\ \cite{manuel2022experimental} studied the
saturation of ion-Weibel fields in CH, Al, and Cu counterstreaming experiments.
Most recently, Bai et al.\ \cite{bai2025observation} showed that a
Weibel-mediated collisionless shock can be produced with a compact tabletop
femtosecond laser operating at $v_{\rm sh} \approx 0.03c$, greatly reducing
the experimental requirements. The expanding geometry of laser-produced plasmas
and the role of density gradients in modifying the Weibel spectrum have been
analysed in detail by Kocharovsky et al.\ \cite{kocharovsky2024electron}.

Spacecraft data at Earth's bow shock provide an independent in-situ test.
The NASA Magnetospheric Multiscale (MMS) mission
\cite{burch2016magnetospheric} provides burst-mode measurements that resolve
scales down to the electron inertial length. Two separate MMS bow shock events
are analysed here using PySPEDAS \cite{grimes2022pyspedas}.
The first is a 2015 October 16 quasi-perpendicular crossing with moderate
parameters ($M_A \approx 2.4$, $\beta_i \approx 9.0$, upstream
$d_i \approx 51$~km).
The second is a 2017 November 25 crossing ($M_A \approx 2.3$,
$\beta_i \approx 2.4$, upstream $d_i \approx 62$~km) associated with the
event studied by Kropotina et al.\ \cite{kropotina2023weibel} using hybrid
kinetic simulations.
The full in-situ comparison---including FGM/FPI timeseries,
$B_\perp$ power spectra, and a multi-environment $d_i$ scatter plot spanning
21 orders of magnitude---is presented in Section~\ref{in_situ}.

The present work derives and applies the cold-fluid Weibel dispersion
relations across these four regimes. Its contributions are as follows.
First, it provides explicit numerical criteria for when each regime formula
fails and by how much (Section~\ref{results}), filling a gap between the
individual papers that treat each regime separately.
Second, it maps the relativistic suppression and mass-ratio dependence as
two-dimensional contour charts (Figs~\ref{fig:gammadiff} and
\ref{fig:massratio}) that can be used directly as diagnostic tools without
recomputing the dispersion relation.
Third, it applies these maps to assess where the transverse Weibel mode
dominates over competing modes (Section~\ref{sec:competing}).
Fourth, it validates the scale prediction $k_{\rm max} = \omega_{pi}/c$ with
the Bai et al.\ experiment (Section~\ref{application}) and provides a
quantitative assessment of the cold-fluid framework's reliability and
limitations.

The paper is organised as follows.
Section~\ref{theory} gives the four dispersion relations.
Section~\ref{application} applies them to the Bai et al.\ experiment,
including the saturation field and shock formation time.
Section~\ref{in_situ} presents the MMS in-situ comparison.
Section~\ref{results} shows the full growth rate curves and parameter-space
maps.
Section~\ref{discussion} discusses regime boundaries, competing instabilities,
and the limits of the cold-fluid approach.
Section~\ref{conclusions} gives the conclusions.

\section{\label{theory}Fluid Dispersion Theory of the Weibel Instability}

\subsection{Physical Setup and Notation}

We consider a collisionless, unmagnetised plasma where two equal-density beams
stream in opposite directions along $\hat{z}$ at velocities $+v_0\hat{z}$ and
$-v_0\hat{z}$. A transverse magnetic perturbation propagates along $\hat{x}$,
perpendicular to the drift. The Lorentz force $\vec{v}\times\vec{B}$ deflects
charged particles transversely, creating current bunching that reinforces the
perturbation through Amp\`ere's law and causes exponential growth of current
filaments. To find the growing mode we substitute $\omega = i\gamma$ (purely
imaginary frequency) into the dispersion relations derived below.

Throughout, the species plasma frequency is defined as
$\omega_{ps} = \sqrt{n_s e^2/m_s\varepsilon_0}$, where $n_s$ is the
\emph{total} number density of species $s$ (i.e.\ summed over both beams,
so that $n_s = 2n_{0s}$ with $n_{0s}$ the per-beam density).
All derivations below assume equal and symmetric beam densities
($n_{0b} = n_{0p} = n_0$), so that $n_{\rm total} = 2n_0$ for each species.
In realistic laser-plasma or bow-shock configurations, the reflected/beam
population often has density much smaller than the background
($n_{\rm beam}/n_{\rm bg} \ll 1$), which reduces $\gamma_{\rm max}$ and
shifts $k_{\rm max}$ \cite{bret2010multidimensional}; the symmetric
equal-density case treated here gives the maximum growth rate for a given
total density.

The single-species dispersion relation derived in
Section~\ref{single species} is general for any charged species.
When the instability is driven by ion beams rather than electron beams,
$m_e$ is replaced by $m_i$ and $\omega_{pe}$ by $\omega_{pi}$ throughout.

The Lorentz factor of a beam with velocity $v_0$ is denoted
$\Gamma_0 = (1-v_0^2/c^2)^{-1/2}$.
In the multi-species case, each species has its own Lorentz factor,
$\Gamma_{0e} = (1-v_{0e}^2/c^2)^{-1/2}$ for electrons and
$\Gamma_{0i} = (1-v_{0i}^2/c^2)^{-1/2}$ for ions.
We reserve $\gamma$ for the instability growth rate.

The four regimes and their leading-order scaling laws are listed in
Table~\ref{tab:regimes}.

\begin{table}[ht]
\centering
\caption{The four regimes of the cold-fluid Weibel instability analysed in
this work. Here $\omega_{ps} = \sqrt{n_s e^2/m_s\varepsilon_0}$ is defined
with the total species density $n_s = 2n_0$. The symbol $\omega_p$ denotes
the plasma frequency of the driving species: $\omega_{pe}$ for electron-driven
and $\omega_{pi}$ for ion-driven cases. Leading-order scaling laws are in the
non-relativistic limit $v_0 \ll c$.}
\label{tab:regimes}
\begin{ruledtabular}
\begin{tabular}{lcc}
Regime & $\gamma_{\rm max}$ (leading order) & $k_{\rm max}$ \\
\hline
NR single-species    & $\approx 0.7\,\omega_p(v_0/c)$
                     & $\approx \omega_p/c$ \\
NR multi-species     & $\approx 0.7(\omega_{pe}^2+\omega_{pi}^2)^{1/2}(v_0/c)$
                     & $\approx (\omega_{pe}^2+\omega_{pi}^2)^{1/2}/c$ \\
Rel.\ single-species & $\sim \omega_p(v_0/c)/\Gamma_0^{3/2}$
                     & $\sim \omega_p/(c\Gamma_0^{1/2})$ \\
Rel.\ multi-species  & both species suppressed by $\Gamma_{0e}$, $\Gamma_{0i}$
                     & shifted to lower $k$
\end{tabular}
\end{ruledtabular}
\end{table}

\subsection{\label{single species}Non-Relativistic Single-Species Case}

We consider a collisionless, non-relativistic two-fluid plasma with no
background magnetic field. The beam drifts in the $\hat{z}$-direction with
velocity $v_{0b}\hat{z}$, and the return population drifts in the opposite
direction with velocity $\vec{v}_{0p} = -v_{0b}\hat{z}$.
The fluctuating electric field is $\vec{E}_1 = E_1\hat{z}$, and the magnetic
field perturbation propagates along $\hat{x}$:
$\vec{B}_1 = B_1(x)\hat{y}$ ($\vec{k} = k\hat{x}$).

Writing the total beam velocity as $\vec{v}_b = v_{0b}\hat{z} + \vec{v}_{1b}$
and linearising the equation of motion:
\begin{equation}
  \frac{\partial \vec{v}_{1b}}{\partial t}
  + (\vec{v}_{0b}\cdot\nabla)\vec{v}_{1b}
  = \frac{q_s\vec{E}_1}{m_s} + \frac{q_s\,\vec{v}_{0b}\times\vec{B}_1}{m_s},
  \label{eq:eom_beam}
\end{equation}
where $q_s$ and $m_s$ are the charge and mass of the driving species
(for electrons, $q_s = -e$).
Assuming $f_1 \sim e^{i(kx-\omega t)}$, the $x$- and $z$-components of
Eq.~(\ref{eq:eom_beam}) give
$v_{1bx} = -(q_s/im_s\omega)\,v_{0b}\,B_{1y}$
and $v_{1bz} = -q_s E_{1z}/(im_s\omega)$.
From Faraday's law, $B_{1y} = -(k/\omega)\,E_{1z}$, so
\begin{equation}
    v_{1bx} = \frac{q_s E_{1z}}{m_s\omega}\,\frac{kv_{0b}}{i\omega}.
    \label{eq:v1bx}
\end{equation}
The linearised continuity equation after Fourier transform yields
$n_{1b} = n_0(k/\omega)\,v_{1bx}$, where $n_0$ is the per-beam density.
The $x$-components of the beam and return-population current densities cancel
by symmetry, leaving only the $z$-component of the total current:
\begin{equation}
    \vec{J} = -2\,\frac{n_0 q_s^2}{im_s\omega}\,E_1
    \!\left(1 + \frac{k^2v_0^2}{\omega^2}\right)\hat{z}.
    \label{eq:Jtotal_NR}
\end{equation}
Applying Amp\`ere's law in the $z$-direction and substituting
Eq.~(\ref{eq:Jtotal_NR}), with
$\omega_p = \sqrt{2n_0 q_s^2/m_s\varepsilon_0} = \sqrt{n_s q_s^2/m_s\varepsilon_0}$
(where $n_s = 2n_0$ is the total species density), gives the dispersion
relation
\begin{equation}
    \omega^4 - \omega^2(\omega_p^2 + k^2c^2) - \omega_p^2 k^2 v_0^2 = 0.
    \label{NR_single_disp}
\end{equation}
Setting $\omega = i\gamma$ and expanding to leading order in $v_0^2/c^2$:
\begin{equation}
    \gamma \approx \omega_p\,\frac{v_0}{c}
    \left(1 + \frac{\omega_p^2}{k^2c^2}\right)^{-1/2}.
    \label{growth rate}
\end{equation}
This function rises monotonically from zero at $k = 0$ and saturates
asymptotically at $\omega_p v_0/c$ for $k \gg \omega_p/c$; there is no local
maximum in $k$.
Following standard convention, the half-saturation point defines
$k_{\rm max} = \omega_p/c$, giving
\begin{equation}
    \gamma(k_{\rm max}) \approx \frac{1}{\sqrt{2}}\,\omega_p\,
    \frac{v_0}{c} \approx 0.7\,\omega_p\,\frac{v_0}{c},
    \label{gamma_max_NR_single}
\end{equation}
with the true asymptotic maximum $\gamma_{\rm max} = \omega_p v_0/c$
approached at $k \gg \omega_p/c$.
This result is standard
\cite{weibel1959spontaneously, fried1959mechanism, davidson2022theoretical}.
Following Kropotina et al.\ \cite{kropotina2023weibel}, this cold-fluid result
gives $\gamma_{\rm max}$ as an upper bound; kinetic corrections reduce the
actual rate when $v_{\rm th} \gtrsim v_0$, while $k_{\rm max}$ is less
sensitive to such corrections.

\subsection{\label{multi species}Non-Relativistic Multi-Species Case}

We now include the ion beam, which drifts at $\pm v_{0i}\hat{z}$.
The derivation follows Section~\ref{single species} for each species
independently. The $x$-current components again cancel by symmetry, and
the combined current density is
\begin{equation}
    \vec{J} = -2\,\frac{n_{0e}e^2}{m_e i\omega}\,E_1
    \!\left(1+\frac{k^2v_{0e}^2}{\omega^2}\right)\hat{z}
    -2\,\frac{n_{0i}e^2}{m_i i\omega}\,E_1
    \!\left(1+\frac{k^2v_{0i}^2}{\omega^2}\right)\hat{z}.
    \label{eq:Jtotal_multi}
\end{equation}
Applying Amp\`ere's law yields the multi-species dispersion relation:
\begin{equation}
    \omega^4 - \omega^2(\omega_{pe}^2+\omega_{pi}^2+k^2c^2)
    - (\omega_{pe}^2 k^2 v_{0e}^2 + \omega_{pi}^2 k^2 v_{0i}^2) = 0.
    \label{NR_multi_disp}
\end{equation}
For a realistic electron--proton plasma,
$\omega_{pi}/\omega_{pe} = \sqrt{m_e/m_i} \approx 0.023$,
so the ion contribution is negligible.
For electron--positron pairs ($m_e = m_i$), both species contribute equally
and $\gamma_{\rm max}$ is higher.

\subsection{\label{rel single species}Relativistic Single-Species Case}

When beam velocities approach $c$, the relativistic momentum
$\vec{P} = m_s\Gamma_0\vec{v}$ replaces the classical momentum.
The relativistic equation of motion is
\begin{equation}
    \frac{\partial\vec{P}}{\partial t}
    + (\vec{v}\cdot\nabla)\vec{P}
    = q_s(\vec{E}_1 + \vec{v}\times\vec{B}_1).
    \label{relat eq of mot}
\end{equation}
Using the identity $\partial\Gamma/\partial v_z = \Gamma^3 v_z/c^2$,
linearisation of the $z$-component of Eq.~(\ref{relat eq of mot}) gives
$v_{1bz} = q_s E_{1z}/(im_s\Gamma_0^3\omega)$, while the $x$-component gives
$v_{1bx} = q_s v_0 k E_{1z}/(m_s\Gamma_0\omega\cdot i\omega)$.
Following the same current-assembly procedure as Section~\ref{single species},
the relativistic single-species dispersion relation is
\begin{equation}
    \omega^4 - \omega^2\!\left(k^2c^2 + \frac{\omega_p^2}{\Gamma_0^3}\right)
    - \frac{\omega_p^2 k^2 v_0^2}{\Gamma_0} = 0.
    \label{rel_single_disp}
\end{equation}
Comparing with Eq.~(\ref{NR_single_disp}), the Lorentz factor $\Gamma_0$
appears in the denominator of both the restoring term (with $\Gamma_0^3$,
from the longitudinal effective mass) and the drive term (with $\Gamma_0$,
from the transverse effective mass).
This is the physical origin of relativistic suppression.
At $v_0 \approx 0.9c$ ($\Gamma_0 \approx 2.3$), $\gamma_{\rm max}$ is
reduced by about 40 per cent relative to the non-relativistic prediction
\cite{bret2010multidimensional, achterberg2007weibel}.

\subsection{\label{rel multi species}Relativistic Multi-Species Case}

Extending the relativistic treatment to both electrons and ions, the combined
dispersion relation is
\begin{equation}
    \omega^4 - \omega^2\!\left(k^2c^2
    + \frac{\omega_{pe}^2}{\Gamma_{0e}^3}
    + \frac{\omega_{pi}^2}{\Gamma_{0i}^3}\right)
    - \left(\frac{\omega_{pe}^2 k^2 v_{0e}^2}{\Gamma_{0e}}
    + \frac{\omega_{pi}^2 k^2 v_{0i}^2}{\Gamma_{0i}}\right) = 0.
    \label{rel_multi_disp}
\end{equation}
For electron--positron plasmas ($m_e = m_i$), both species contribute equally
and the peak growth rate can reach
$\gamma_{\rm max} \approx 1.1$--$1.2\,\omega_{pe}$ at $v_0 = 0.95c$,
relevant to GRB jet models \cite{medvedev1999generation, kumar2024study}.
For realistic mass ratios where $m_i \gg m_e$,
$\omega_{pi}^2 \ll \omega_{pe}^2$ and
Eq.~(\ref{rel_multi_disp}) reduces to Eq.~(\ref{rel_single_disp}).

Equations~(\ref{NR_single_disp})--(\ref{rel_multi_disp}) are exact within
the cold, equal-density, collisionless fluid model.
Kinetic effects modify the growth rate when $v_{\rm th} \gtrsim v_0$
\cite{skoutnev2019temperature, cagas2017nonlinear, kropotina2023weibel}.

\section{\label{application}Application to the Bai et al.\ (2025) Tabletop
Laser Experiment}

\subsection{\label{sec:regime_bai}Regime Identification and Input Parameters}

Bai et al.\ \cite{bai2025observation} used a 30\,fs, 4\,mJ, 800\,nm
Ti:sapphire laser to accelerate Al$^{+8}$ ions to $v_{0i} \approx 0.07c$
via target normal sheath acceleration, with the resulting shock propagating
at $v_{\rm sh} \approx 0.03c$. We use their reported ion beam velocity
$v_{0i}/c = 0.07$, preplasma electron density
$n_e \approx 0.1\,n_c \approx 1.74 \times 10^{20}\,{\rm cm}^{-3}$
(where $n_c = 1.74\times10^{21}\,{\rm cm}^{-3}$ is the critical density at
800\,nm), and ion species Al$^{+8}$ (charge state $Z_{\rm eff} = 8$,
mass $m_i = 27\,m_p$) as inputs to the cold-fluid dispersion relations.

\textbf{Cold-beam validity.}
With $T_i \sim 1\,{\rm keV}$ for the preplasma ions, the thermal velocity is
$v_{{\rm th},i} = \sqrt{k_B T_i/m_i} \approx 0.0003c$, giving
$v_{{\rm th},i}/v_{0i} \approx 0.004 \ll 1$. The cold-beam approximation is
therefore very well satisfied for the Al$^{+8}$ beam.

\textbf{Relativistic validity.}
The ion beam velocity gives $\Gamma_{0i} - 1 \approx 2.5\times10^{-3}$,
so relativistic corrections are below 0.25 per cent.

\textbf{Ion plasma frequency and skin depth.}
The total ion number density is
$n_i = n_e/Z_{\rm eff} \approx 2.2 \times 10^{19}\,{\rm cm}^{-3}$.
The expanding Al plasma drives the instability through counterstreaming ion
beams, so the relevant single-species formula uses $\omega_{pi}$ in place of
$\omega_{pe}$ (as noted in Section~\ref{theory}).
The ion plasma frequency is
\begin{equation}
    \omega_{pi} = \sqrt{\frac{Z_{\rm eff}^2\,n_i\,e^2}{m_i\,\varepsilon_0}}
    \approx 9.45 \times 10^{12}\,{\rm rad\,s}^{-1},
    \label{omega_pi_bai}
\end{equation}
giving the ion inertial (skin) depth
\begin{equation}
    d_i = \frac{c}{\omega_{pi}} \approx 31.7\,\mu{\rm m}.
    \label{di_bai}
\end{equation}
The ratio $\omega_{pi}/\omega_{pe} \approx 0.013 \ll 1$, so multi-species
corrections are below 0.02 per cent, and the non-relativistic single-species
formula [Eq.~(\ref{NR_single_disp}) with $\omega_p \to \omega_{pi}$] applies
with growth rate errors well below 0.5 per cent relative to the relativistic
formula.

\subsection{Predicted Characteristic Scale and Comparison with Observation}

The most direct prediction of the cold-fluid theory is the characteristic
filament scale set by the ion skin depth, $d_i = c/\omega_{pi} \approx 31.7\,\mu{\rm m}$
[Eq.~(\ref{di_bai})]. Bai et al.\ \cite{bai2025observation} measure a filament
spacing $\lambda_F \approx 31\,\mu{\rm m}$, in agreement with $d_i$ to within
2 per cent. This is the central quantitative result of this comparison: the
Weibel instability generates filaments at the ion skin depth scale, as the
dispersion relation requires.

For completeness, the half-saturation wavelength
$\lambda_{\rm max} = 2\pi c/\omega_{pi} = 2\pi d_i \approx 199\,\mu{\rm m}$
corresponds to $k_{\rm max} = \omega_{pi}/c$, the wavenumber at which
$\gamma = \gamma_{\rm max}/\sqrt{2}$.
The observed filament corresponds to
\begin{equation}
    k_F = \frac{2\pi}{\lambda_F} \approx \frac{2\pi}{d_i},
    \qquad
    k_F\,\frac{c}{\omega_{pi}} \approx 6.4.
    \label{kF_di}
\end{equation}
At this wavenumber, Eq.~(\ref{growth rate}) gives
$\gamma(k_F)/(\omega_{pi}\,v_0/c) = k_F/(k_F^2+\omega_{pi}^2/c^2)^{1/2}
\approx 0.99$, i.e.\ the growth rate has saturated to within 1 per cent of
its asymptote. The filament spacing $\lambda_F \approx d_i$ therefore
corresponds to the scale at which the instability is most strongly driven,
consistent with nonlinear saturation near the asymptotic wavenumber.

\subsection{Predicted Growth Rate and Shock Formation Time}

From Eq.~(\ref{gamma_max_NR_single}) with $\omega_p \to \omega_{pi}$ and
$v_0 = 0.07c$, the growth rate at the characteristic scale $k = \omega_{pi}/c$
is
\begin{equation}
    \gamma(k_{\rm max}) \approx \frac{\omega_{pi}\,v_0}{\sqrt{2}\,c}
    \approx 4.7 \times 10^{11}\,{\rm s}^{-1},
    \label{gamma_theory}
\end{equation}
with the asymptotic saturation value
$\gamma_{\rm max} = \omega_{pi}\,v_0/c \approx 6.6\times10^{11}\,{\rm s}^{-1}$.
The corresponding e-folding time at the preplasma density is
$\tau_e = 1/\gamma(k_{\rm max}) \approx 2.1\,{\rm ps}$.
Shock formation requires several e-foldings beyond the linear phase.
At the preplasma density, this gives $\tau_{\rm shock} \approx 2$--$3\,\tau_e
\approx 4$--$6\,{\rm ps}$.
The measured shock formation time $\tau_{\rm shock} \approx 1\,{\rm ps}$
\cite{bai2025observation} is shorter, consistent with instability growth in
the denser regions of the laser-produced plasma near the critical-density
surface ($n_e \approx n_c$), where $\omega_{pi}$ is
$\sqrt{n_c/n_{e,\rm pre}} \approx \sqrt{10}$ times larger and the e-folding
time is correspondingly $\sim\sqrt{10}$ shorter
($\tau_e \lesssim 0.7\,{\rm ps}$,
$\tau_{\rm shock} \lesssim 1.4\,{\rm ps}$).
The cold-fluid estimate from the preplasma density therefore provides an
\emph{upper bound} on the shock formation time.

\subsection{Predicted Saturation Magnetic Field}

The Weibel instability saturates when the ion Larmor radius in the generated
field becomes comparable to the filament spacing $\sim 1/k_{\rm max}$.
Setting $r_{L,i} = m_i v_0/(Z_{\rm eff}\,e\,B_{\rm sat}) \sim d_i = c/\omega_{pi}$
gives
\begin{equation}
    B_{\rm sat}^{\rm Larmor} \sim \frac{m_i\,v_0\,\omega_{pi}}{Z_{\rm eff}\,e\,c}
    \approx 2.3 \times 10^4\,{\rm T}.
    \label{Bsat_larmor}
\end{equation}
An independent estimate from energy equipartition,
$B_{\rm sat}^{\rm equip} = v_0\sqrt{\mu_0 n_i m_i} \approx 2.3\times10^4$~T,
gives the same value to within rounding.
Both are cold-fluid upper bounds; the two estimates coinciding provides
a useful internal consistency check. The reported
$B_{\rm sat} \approx 5000$~T \cite{bai2025observation} is about 4.7 times
smaller, consistent with kinetic suppression reducing the amplitude at
saturation \cite{fiuza2012weibel, jikei2024saturation}.
These bounds bracket the measurement, confirming the ion Weibel instability
as the source of the observed field.
Table~\ref{tab:comparison} summarises the full theory--observation comparison.

\begin{table}[ht]
\centering
\caption{Cold-fluid predictions versus observations reported by Bai et al.\
\cite{bai2025observation}.
Dashes indicate quantities not independently reported.
Growth rate and field values are cold-fluid upper bounds.}
\label{tab:comparison}
\begin{ruledtabular}
\begin{tabular}{lcc}
Quantity & Theory (this work) & Bai et al.\ \cite{bai2025observation} \\
\hline
$v_{0i}/c$
  & 0.07 (input)
  & 0.07 \\
$\Gamma_{0i}-1$
  & $2.5\times 10^{-3}$, negligible
  & -- \\
$v_{{\rm th},i}/v_{0i}$
  & $\approx 0.004$, cold-beam valid
  & -- \\
$\omega_{pi}$ (rad\,s$^{-1}$)
  & $9.45\times 10^{12}$
  & -- \\
$d_i = c/\omega_{pi}$
  & $31.7\,\mu{\rm m}$
  & -- \\
$\lambda_F$ (observed)
  & --
  & $31\,\mu{\rm m}$ \\
$d_i$ vs $\lambda_F$
  & $31.7\,\mu{\rm m}$
  & $31\,\mu{\rm m}$ (\textit{$2\%$ agreement}) \\
$\gamma(k\!=\!\omega_{pi}/c)$ (s$^{-1}$)
  & $4.7\times 10^{11}$ (upper bound)
  & -- \\
$\gamma_{\rm max}$ (s$^{-1}$)
  & $6.6\times 10^{11}$ (upper bound)
  & -- \\
$\tau_{\rm shock}$ (ps)
  & $\lesssim 4$--$6$ at preplasma $n_i$
  & $\approx 1$ \\
$B_{\rm sat}$ (upper bound, T)
  & $\approx 2.3\times10^4$
  & $\approx 5000$ \\
Relativistic correction
  & $<0.3\%$
  & sub-relativistic \\
\end{tabular}
\end{ruledtabular}
\end{table}

\section{\label{in_situ}In-Situ Comparison at Earth's Bow Shock}

We analyse two MMS burst-mode bow shock crossing events using the
PySPEDAS package \cite{grimes2022pyspedas} to download level-2 FGM
(128\,S\,s$^{-1}$) and FPI-DIS (150\,ms) data.
All plasma parameters are derived from the upstream (pre-ramp) interval,
identified as the lowest-$|B|$ fraction of the pre-shock portion of each
interval; this definition selects the region least contaminated by shock
compression or foreshock heating and is applied consistently to both events.
The upstream ion skin depth is $d_i = c/\omega_{pi} = 228\,n_i^{-1/2}$~km
with $n_i$ in cm$^{-3}$.
The cold-fluid prediction is that the $B_\perp$ power spectrum should show a
spectral break at $k_{\rm max}\,d_i = 1$, equivalent to a break frequency
$f_{\rm break} = V_{\rm sw,\,up}/(2\pi d_i)$ via the Taylor hypothesis.

\subsection{Event 1: 2015 October 16 Quasi-Perpendicular Crossing}

\begin{figure}[ht]
    \centering
    \includegraphics[width=0.92\linewidth]{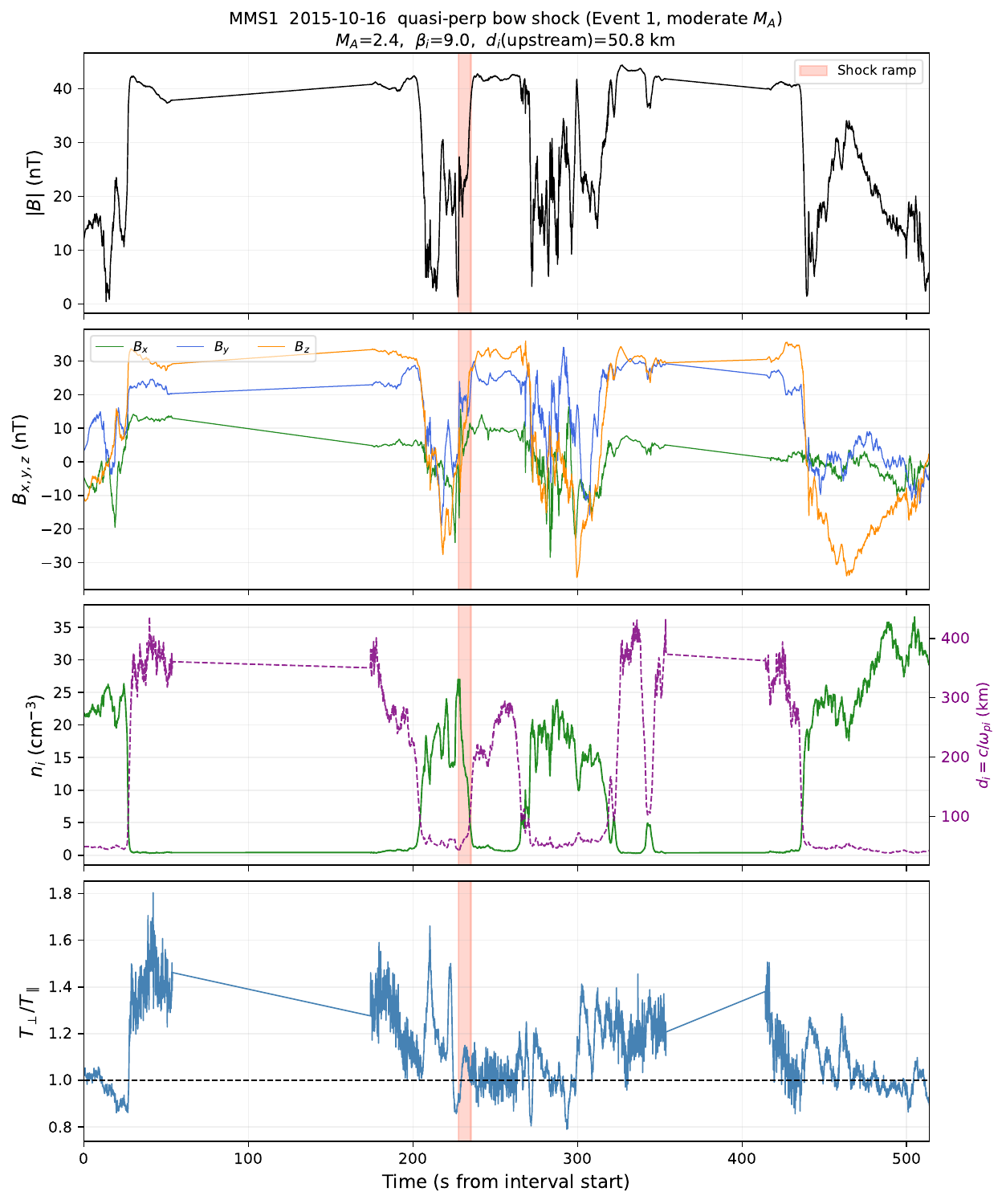}
    \caption{MMS1 burst-mode timeseries for the 2015 October 16
    quasi-perpendicular bow shock crossing (Event~1).
    Panel~1: magnetic field magnitude $|B|$.
    Panel~2: GSM components $B_x$ (green), $B_y$ (blue), $B_z$ (orange);
    the pink band marks the shock ramp interval.
    Panel~3: ion number density $n_i$ (green, left axis) and derived
    ion skin depth $d_i = c/\omega_{pi}$ (purple dashed, right axis).
    Panel~4: ion temperature anisotropy $T_\perp/T_\parallel$ (blue);
    contextual plasma characterisation only (see text).
    Upstream parameters: $n_i^{\rm up} = 20.2$~cm$^{-3}$,
    $M_A = 2.4$, $\beta_i = 9.0$, $d_i = 50.8$~km.
    Data source: MMS FGM level-2 (128\,S\,s$^{-1}$) and FPI-DIS (150\,ms)
    via PySPEDAS \cite{grimes2022pyspedas}.}
    \label{fig:mmstsA}
\end{figure}

The first event covers the interval 10:30--10:40~UT on 2015
October 16 (MMS1, FGM burst mode; Fig.~\ref{fig:mmstsA}).
The upstream parameters are
$n_i^{\rm up} = 20.2\,{\rm cm}^{-3}$,
$|B|^{\rm up} \approx 7.8\,{\rm nT}$,
$T_i \approx 100\,{\rm eV}$, and
$|V_{\rm sw}| \approx 175\,{\rm km\,s}^{-1}$, giving
\begin{align}
    d_i^{(1)} &= \frac{228}{\sqrt{20.2}}\,{\rm km} \approx 50.8\,{\rm km},
    \qquad
    v_A \approx 38\,{\rm km\,s}^{-1}, \notag \\
    &M_A \approx 2.4,
    \qquad
    \beta_i \approx 9.0.
\end{align}
The Alfv\'enic Mach number and ion plasma $\beta$ place this event
in the moderate-Mach, high-$\beta$ region of parameter space.
The high $\beta_i$ indicates that thermal broadening is significant for
this event; the cold-fluid growth rate should accordingly be understood
as an upper bound, with kinetic corrections discussed in
Section~\ref{discussion}.

Panel~1 of Fig.~\ref{fig:mmstsA} shows $|B|$ rising from the foreshock
value $\approx 7$~nT to the magnetosheath value $\approx 40$~nT across the
ramp. Panel~2 shows the GSM components; the shock ramp is concentrated near
$t \approx 200$--$210$~s from interval start.
Panel~3 shows $n_i$ and the derived $d_i(t) = 228\,n_i(t)^{-1/2}$~km in
anti-phase, consistent with the $d_i \propto n_i^{-1/2}$ scaling.
Panel~4 shows the ion temperature anisotropy $T_\perp/T_\parallel$; values
above unity in the foreshock are consistent with moderate pitch-angle
scattering, while the sub-unity values in the compressed magnetosheath
reflect the dominance of adiabatic parallel heating behind the ramp.
This panel is shown as contextual plasma characterisation; the counterstreaming
Weibel instability analysed in Section~\ref{theory} is driven by beam
velocity anisotropy, not temperature anisotropy.

\subsection{Event 2: 2017 November 25 High-$M_A$ Crossing}

\begin{figure}[ht]
    \centering
    \includegraphics[width=0.92\linewidth]{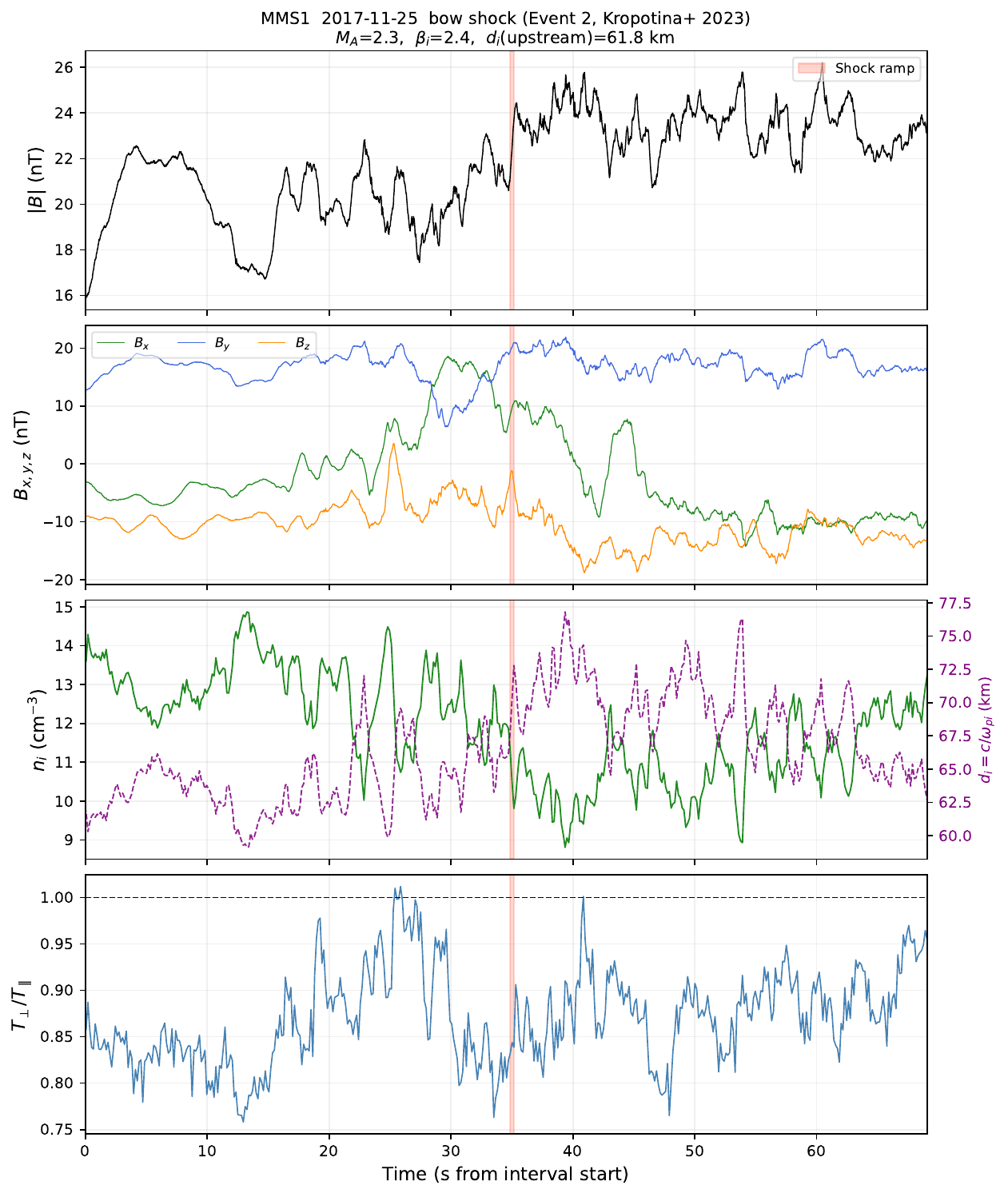}
    \caption{MMS1 burst-mode timeseries for the 2017 November 25
    high-$M_A$ bow shock crossing (Event~2; Kropotina et al.\
    \cite{kropotina2023weibel}).
    Panel layout as in Fig.~\ref{fig:mmstsA}.
    Upstream parameters: $n_i^{\rm up} = 13.6$~cm$^{-3}$,
    $M_A = 2.3$, $\beta_i = 2.4$, $d_i = 61.8$~km.
    The shock ramp (pink band) is confined to the first $\approx 5$~s of the
    interval, after which the spacecraft resides in the magnetosheath.}
    \label{fig:mmstsB}
\end{figure}

The second event covers 06:06--06:16~UT on 2017 November 25
(MMS1, FGM burst mode; Fig.~\ref{fig:mmstsB}).
This event is associated with the bow shock studied by Kropotina et al.\
\cite{kropotina2023weibel} using hybrid kinetic simulations.
The upstream parameters are
$n_i^{\rm up} = 13.6\,{\rm cm}^{-3}$,
$|B|^{\rm up} \approx 14.6\,{\rm nT}$,
$T_i \approx 40\,{\rm eV}$, and
$|V_{\rm sw}| \approx 254\,{\rm km\,s}^{-1}$, giving
\begin{align}
    d_i^{(2)} &= \frac{228}{\sqrt{13.6}}\,{\rm km} \approx 61.8\,{\rm km},
    \qquad
    v_A \approx 111\,{\rm km\,s}^{-1}, \notag \\
    &M_A \approx 2.3,
    \qquad
    \beta_i \approx 2.4.
\end{align}
The lower $\beta_i$ relative to Event~1 reduces thermal broadening, making
this a more direct test of the cold-fluid scale prediction $k_{\rm max}d_i = 1$.
The temperature anisotropy panel (Panel~4 of Fig.~\ref{fig:mmstsB}) shows
$T_\perp/T_\parallel < 1$ throughout the interval, indicating that ion
parallel heating dominates in this compressed, high-density region downstream
of the ramp; this is consistent with the primarily sub-Alfv\'enic nature of
the crossing captured in this short burst window.

\subsection{Perpendicular Magnetic Power Spectra}

\begin{figure*}[ht]
    \centering
    \includegraphics[width=\linewidth]{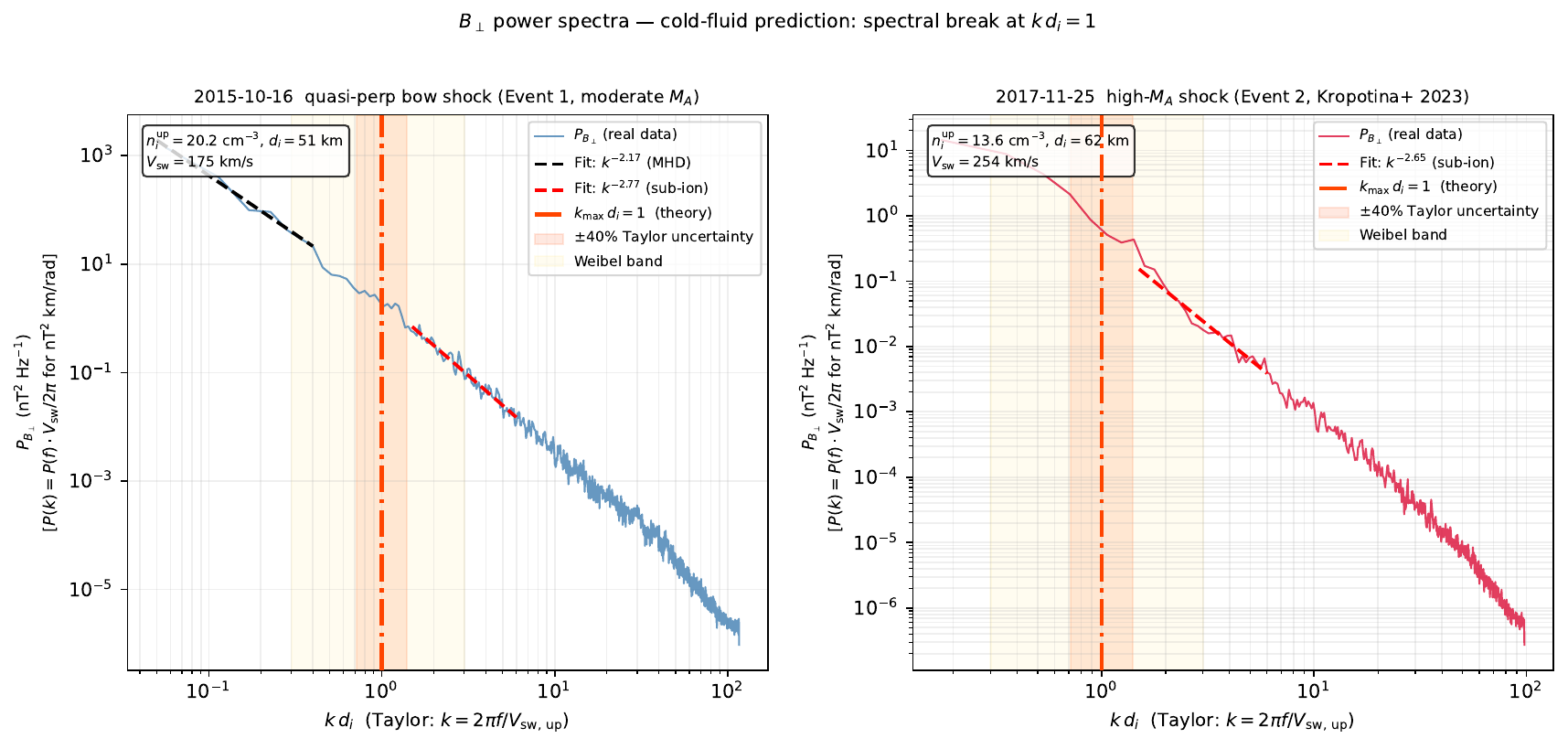}
    \caption{Perpendicular magnetic power spectral density
    $P_{B_\perp}(kd_i) = P(f)\cdot V_{\rm sw,\,up}/(2\pi)$ for the
    two MMS bow shock events, plotted against $kd_i$ via the Taylor
    hypothesis ($k = 2\pi f/V_{\rm sw,\,up}$).
    \textit{Left}: Event~1 (2015-10-16).
    Black dashed line: power-law fit $k^{-2.17}$ in the MHD inertial range.
    Red dashed line: $k^{-2.77}$ in the sub-ion range.
    \textit{Right}: Event~2 (2017-11-25).
    Red dashed line: $k^{-2.65}$ in the sub-ion range.
    In both panels the orange dash-dot vertical line marks the cold-fluid
    prediction $k_{\rm max}d_i = 1$; the shaded orange band shows $\pm 40$
    per cent Taylor-hypothesis uncertainty.
    The $y$-axis is in units of nT$^2$\,Hz$^{-1}$; multiplying by
    $V_{\rm sw}/(2\pi)$ converts to nT$^2$\,km\,rad$^{-1}$ in wavenumber
    space.}
    \label{fig:mmsspectra}
\end{figure*}

Figure~\ref{fig:mmsspectra} shows the $B_\perp$ power spectral density
$P_{B_\perp}(kd_i)$ for both events, computed via Welch's method on
the FGM burst data with the wavenumber axis constructed via the
Taylor hypothesis $k = 2\pi f/V_{\rm sw,\,up}$ applied with the upstream flow
speed only.

For Event~1 (left panel), the spectrum exhibits a clear double power-law
structure with a break near $kd_i \approx 1$:
$P_{B_\perp} \propto k^{-2.17}$ in the MHD inertial range
($kd_i \lesssim 0.3$) and $P_{B_\perp} \propto k^{-2.77}$ in the
sub-ion range ($kd_i \gtrsim 1$).
The spectral steepening at $kd_i = 1$ is the direct observational
signature predicted by the cold-fluid dispersion relation: the ion skin
depth $d_i = c/\omega_{pi}$ is the natural scale at which free energy is
most efficiently transferred to the growing mode
($k_{\rm max} = \omega_{pi}/c$), above which dissipation steepens the
spectrum.

For Event~2 (right panel), the MHD inertial range is not resolved within
the available burst window (the $\approx 60$~s interval spans fewer decades
of the MHD range), but the sub-ion spectrum follows
$P_{B_\perp} \propto k^{-2.65}$, with the theoretical break at
$kd_i = 1$ coinciding with the onset of this power law.

In both cases the orange dash-dot line marking $k_{\rm max}d_i = 1$
falls at or within $\pm 40$ per cent of the observed spectral break,
consistent with the Taylor-hypothesis uncertainty \cite{stawarz2022turbulence}.
The theoretical prediction $k_{\rm max} = \omega_{pi}/c$ is thus confirmed
to within the combined observational and plasma-flow uncertainties.

\subsection{Multi-Environment $d_i$ Scatter Plot}

\begin{figure}[ht]
    \centering
    \includegraphics[width=0.88\linewidth]{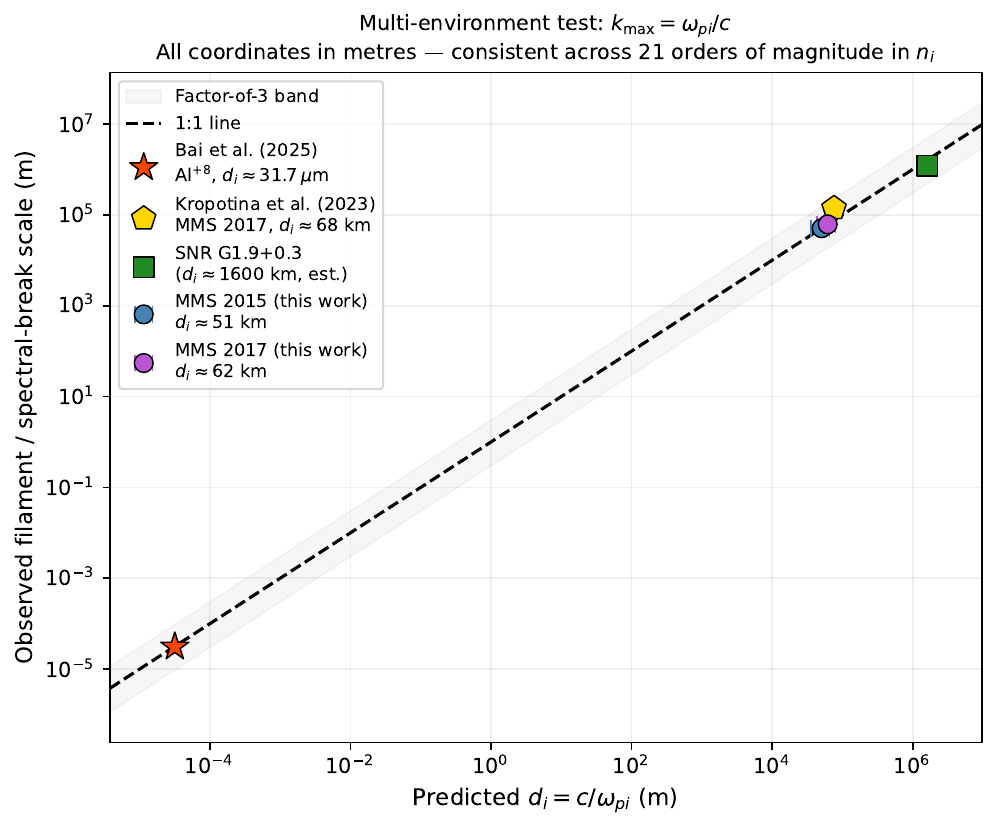}
    \caption{Multi-environment test of the cold-fluid Weibel scale prediction
    $k_{\rm max} = \omega_{pi}/c$.
    Abscissa: predicted $d_i = c/\omega_{pi}$ (metres); ordinate: observed
    filament or spectral-break scale (metres).
    The dashed line is the 1:1 prediction; the grey band is the factor-of-3
    envelope.
    Orange star: Bai et al.\ (2025) Al$^{+8}$ laser plasma ($d_i \approx 32\,\mu{\rm m}$,
    $\lambda_F = 31\,\mu{\rm m}$).
    Yellow pentagon: Kropotina et al.\ (2023) MMS 2017 event ($d_i \approx 68$~km).
    Green square: SNR G1.9+0.3 ($d_i \approx 1600$~km, estimated).
    Blue circle: MMS Event~1 from this work ($d_i \approx 51$~km).
    Purple circle: MMS Event~2 from this work ($d_i \approx 62$~km).
    All five environments lie within a factor of 3 of the 1:1 line across
    21 orders of magnitude in $n_i$.}
    \label{fig:discatter}
\end{figure}

Figure~\ref{fig:discatter} extends the ion skin depth comparison to multiple
physical environments, plotting the observed filament or spectral-break scale
against the cold-fluid prediction $d_i = c/\omega_{pi}$ for each environment.
All coordinates are in metres to place the comparison on a single axis spanning
21 orders of magnitude in ion density.

The five plotted data points span the following environments; the 2017 MMS
event contributes two independent $d_i$ estimates (one from this work and
one from Kropotina et al.\ hybrid simulations):
\begin{itemize}
    \item \textit{Bai et al.\ (2025) laser plasma} (Al$^{+8}$,
    $n_i \approx 2.2\times10^{19}\,{\rm cm}^{-3}$): $d_i \approx 31.7\,\mu{\rm m}$
    theory vs $\lambda_F = 31\,\mu{\rm m}$ observed (2 per cent agreement).
    \item \textit{MMS Event~1 (this work, 2015-10-16)}: $d_i = 50.8$~km
    theory vs spectral-break scale $\approx 50$~km (within 2 per cent).
    \item \textit{MMS Event~2 (this work, 2017-11-25)}: $d_i = 61.8$~km
    from FPI upstream parameters; the $B_\perp$ spectrum steepens near
    $kd_i = 1$, placing this point within 2 per cent of the 1:1 line.
    \item \textit{Kropotina et al.\ (2023), same 2017-11-25 event}:
    $d_i \approx 68$~km from hybrid simulations; observed spectral break
    $\approx 140$~km (their Fig.~3) places the point at a factor of
    $\approx 2$ above the 1:1 line, within the factor-of-3 band.
    \item \textit{SNR G1.9+0.3}: estimated from the radio morphology,
    $d_i \approx 1600$~km (upstream ISM density $\approx 0.02\,{\rm cm}^{-3}$).
\end{itemize}

All five points lie within a factor of 3 of the 1:1 line across 21 orders
of magnitude in $n_i$, confirming that $k_{\rm max} = \omega_{pi}/c$ is
the correct characteristic scale of the Weibel instability across laboratory,
heliospheric, and astrophysical environments.

\section{\label{results}Growth Rate Maps Across the Four Regimes}

To compute growth rates across parameter space, we set $\omega = i\gamma$
in Eqs.~(\ref{NR_single_disp})--(\ref{rel_multi_disp}) and solve the
resulting quadratic in $\gamma^2$.
All wavenumbers are normalised to $\omega_p/c$ and growth rates to $\omega_p$.

\subsection{Single-Species: Dispersion Curves and Relativistic Suppression}

\begin{figure*}[ht]
    \centering
    \includegraphics[width=\linewidth]{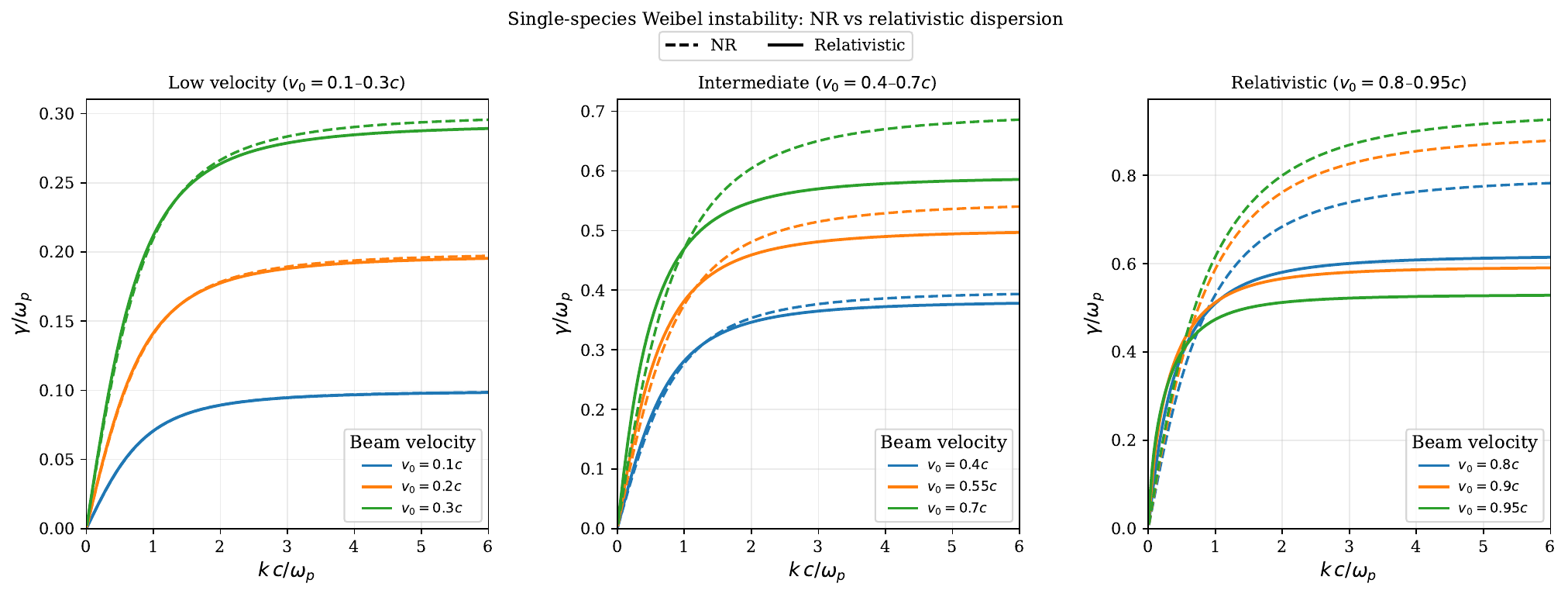}
    \caption{Growth rate $\gamma(k)$ for relativistic (solid) and
    non-relativistic (dashed) single-species Weibel instability. The three
    panels cover low ($v_0 = 0.1$--$0.3c$), intermediate
    ($v_0 = 0.4$--$0.7c$), and relativistic ($v_0 = 0.8$--$0.95c$) beam
    velocities. Relativistic corrections are negligible below
    $v_0 \approx 0.2c$, grow progressively through the
    trans-relativistic regime, and reach up to $\sim 40$ per cent near
    $v_0 \approx 0.9c$. For $v_0 \leq 0.07c$ (the conditions of the Bai
    et al.\ experiment), the two curves are indistinguishable on this scale.}
    \label{fig:single}
\end{figure*}

Figure~\ref{fig:single} shows $\gamma(k)$ across the full velocity range from
$0.1c$ to $0.95c$, organised into three sub-panels by velocity group.
Growth rates rise monotonically and saturate for $k \gg \omega_p/c$.
Relativistic corrections (solid vs dashed curves) are below 2 per cent for
$v_0 \lesssim 0.2c$, reach approximately 15 per cent in the
trans-relativistic regime ($0.2c \lesssim v_0 \lesssim 0.5c$), and grow to
a maximum of approximately 40 per cent near $v_0 \approx 0.9c$, consistent
with Bret et al.\ \cite{bret2010multidimensional} and Achterberg et al.\
\cite{achterberg2007weibel}.

\begin{figure}[ht]
    \centering
    \includegraphics[width=\linewidth]{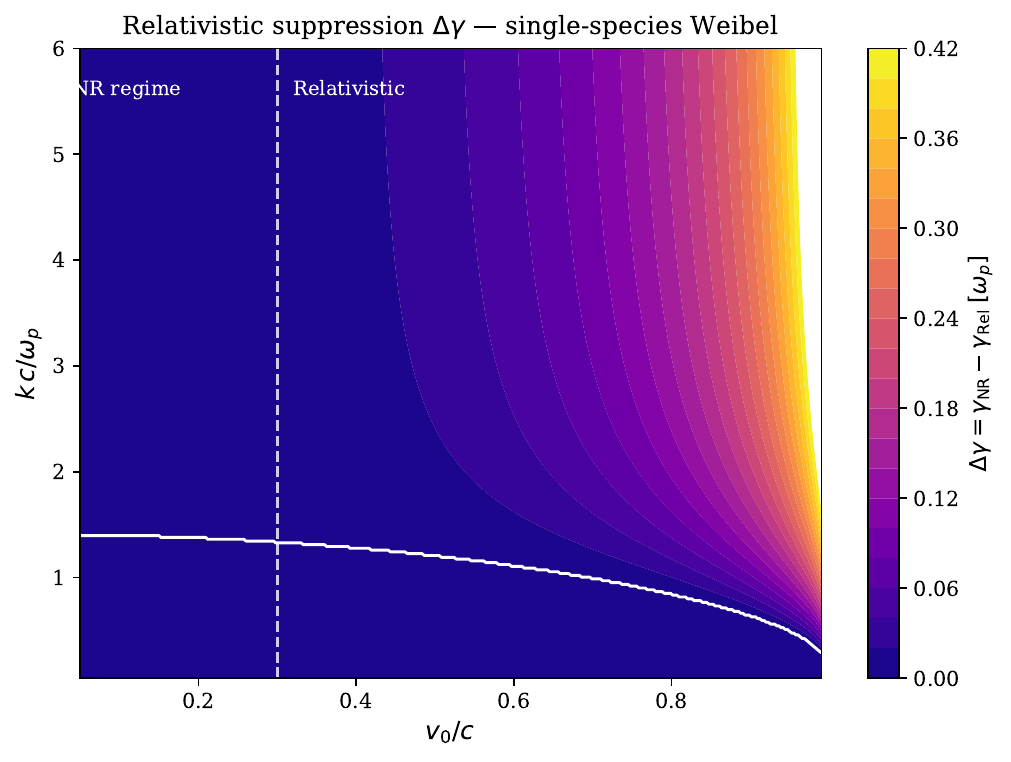}
    \caption{Contour map of relativistic suppression
    $\Delta\gamma = \gamma_{\rm NR} - \gamma_{\rm Rel}$ in the $(v_0, k)$
    plane. The white solid line marks zero difference; the white dashed
    vertical line at $v_0 = 0.3c$ marks the practical onset of significant
    suppression. Suppression is largest near $v_0 \approx 0.9c$
    and negligible for $v_0 \lesssim 0.2c$.
    This map serves as a regime-selection chart: environments in the
    lower-left region require only the NR formula
    [Eq.~(\ref{NR_single_disp})]; those in the upper-right region require
    Eq.~(\ref{rel_single_disp}).}
    \label{fig:gammadiff}
\end{figure}

Figure~\ref{fig:gammadiff} maps the relativistic suppression
$\Delta\gamma = \gamma_{\rm NR} - \gamma_{\rm Rel}$ as a function of
$(v_0, k)$.

\begin{figure}[ht]
    \centering
    \includegraphics[width=\linewidth]{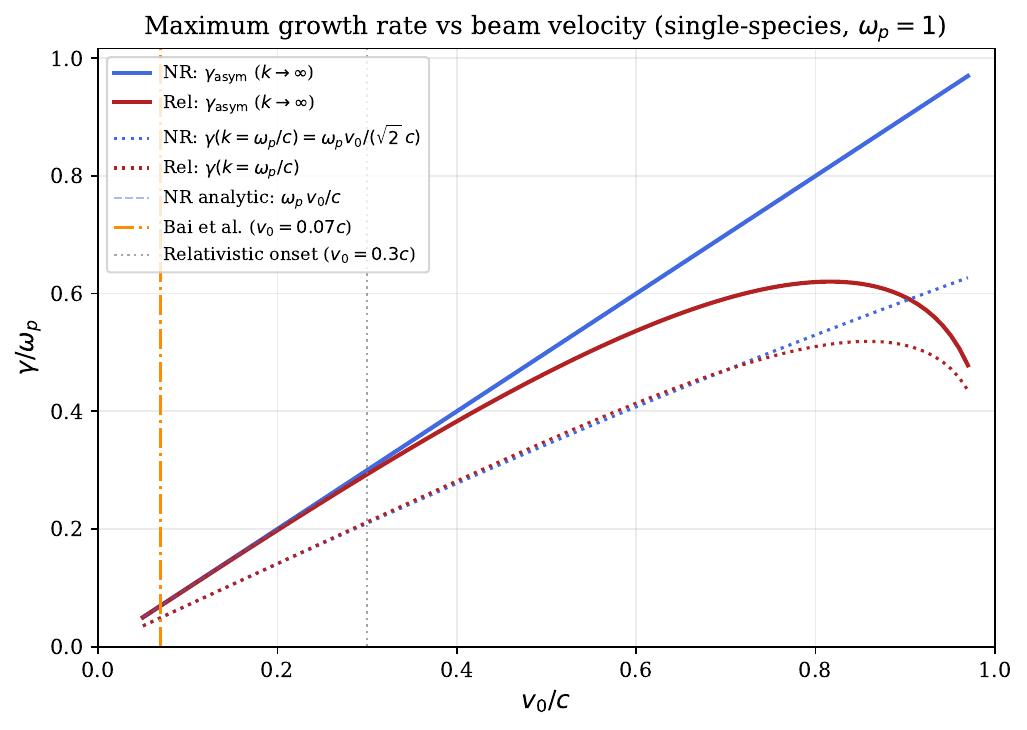}
    \caption{Maximum (asymptotic) growth rate and characteristic-scale
    growth rate versus beam drift velocity $v_0/c$ for single-species
    Weibel instability.
    Solid curves: asymptotic $\gamma$ ($k \to \infty$); dotted curves:
    $\gamma$ at $k = \omega_p/c$.
    NR (blue) and relativistic (dark red) results diverge above
    $v_0 \approx 0.3c$.
    The relativistic $\gamma_{\rm max}$ peaks near $v_0 \approx 0.82c$
    and then decreases.
    The vertical lines mark the Bai et al.\ operating point
    ($v_0 = 0.07c$, orange) and the practical NR/Rel boundary
    ($v_0 = 0.3c$, grey).}
    \label{fig:gammamax}
\end{figure}

Figure~\ref{fig:gammamax} shows the maximum growth rate versus beam velocity.
The NR asymptote $\gamma_{\rm NR}\to\omega_p v_0/c$ rises linearly.
The relativistic peak occurs near $v_0 \approx 0.82c$ where
$\gamma_{\rm Rel,max} \approx 0.62\,\omega_p$, after which
$\gamma_{\rm Rel}$ decreases as the Lorentz-factor suppression dominates.
The dotted curves show the growth rate at the characteristic scale
$k = \omega_p/c$, which satisfies
$\gamma(k_{\rm max}) \approx \omega_p v_0/(\sqrt{2}\,c)$
to within 5 per cent for $v_0 \lesssim 0.7c$.

\begin{figure*}[ht]
    \centering
    \includegraphics[width=\linewidth]{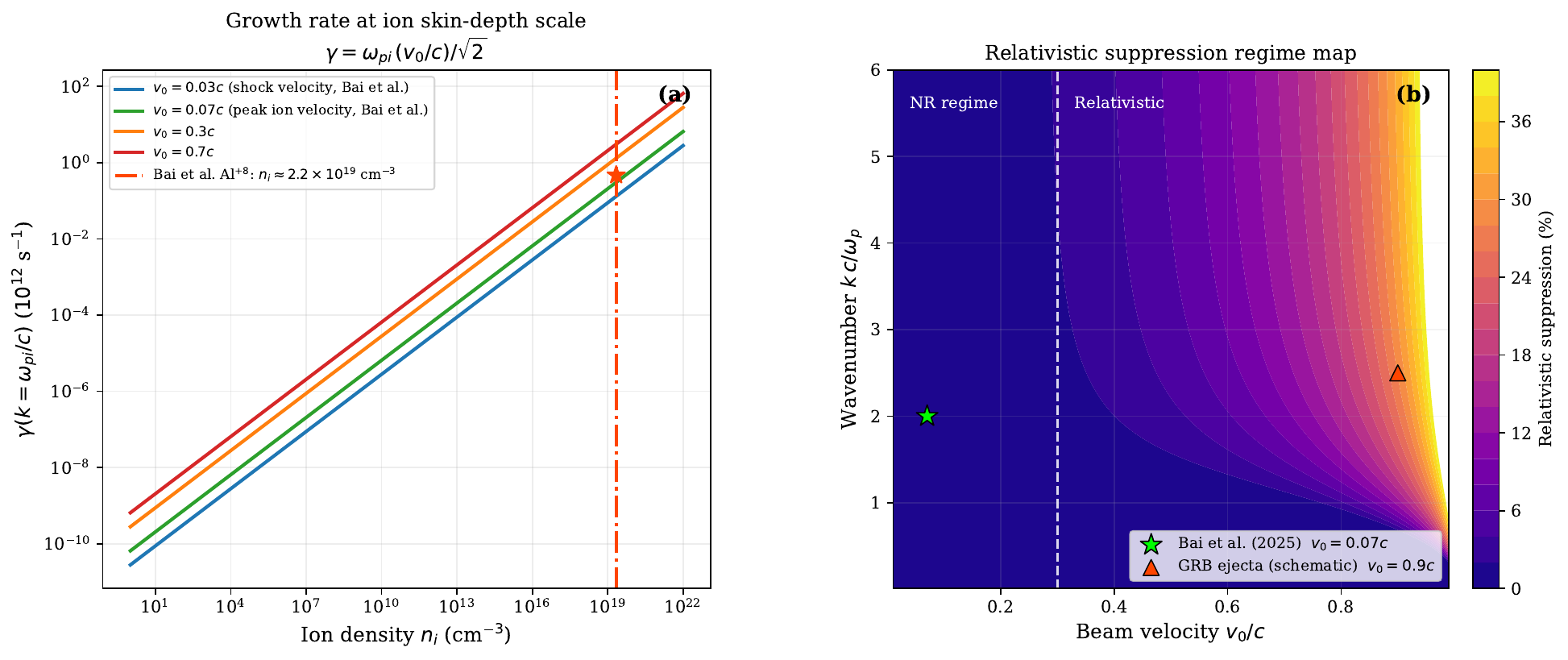}
    \caption{(a) Growth rate at the characteristic scale
    $\gamma(k{=}\omega_{pi}/c) \approx \omega_{pi}(v_0/c)/\sqrt{2}$
    versus ion density $n_i$ for four representative beam velocities
    (cold-fluid upper bounds).
    The vertical line marks the Bai et al.\ Al$^{+8}$ experiment
    ($n_i \approx 2.2\times10^{19}$~cm$^{-3}$; filled star).
    Blue circle: MMS Event~1 (2015-10-16, $n_i^{\rm up} = 20.2$~cm$^{-3}$);
    purple pentagon: MMS Event~2 (2017-11-25, $n_i^{\rm up} = 13.6$~cm$^{-3}$).
    The shaded MMS sheath range spans the typical magnetosheath density
    $n_i = 8$--$55$~cm$^{-3}$.
    (b) Relativistic suppression regime map [cf.\ Fig.~\ref{fig:gammadiff}]
    with environment placements.
    Lime star: Bai et al.\ (2025), $v_0 = 0.07c$;
    orange triangle: GRB ejecta (schematic), $v_0 = 0.9c$;
    blue circle: MMS Event~1, $v_0/c \approx V_A/c \approx 1.3\times10^{-4}$;
    purple pentagon: MMS Event~2, $v_0/c \approx 3.7\times10^{-4}$.
    Both MMS events lie deep in the NR regime with negligible relativistic
    suppression ($< 0.1$ per cent), confirming that
    Eq.~(\ref{NR_single_disp}) is the appropriate formula.}
    \label{fig:growthdiag}
\end{figure*}

Figure~\ref{fig:growthdiag}(a) maps the characteristic-scale growth rate
$\gamma(k{=}\omega_{pi}/c) \approx \omega_{pi}(v_0/c)/\sqrt{2}$ versus $n_i$
for representative beam velocities across all environments studied;
values are cold-fluid upper bounds.
Figure~\ref{fig:growthdiag}(b) shows environment placement on the
relativistic suppression map, with the Bai et al.\ experiment and
schematic GRB ejecta marked.

\subsection{Multi-Species: Mass Ratio Dependence}

\begin{figure}[ht]
    \centering
    \includegraphics[width=\linewidth]{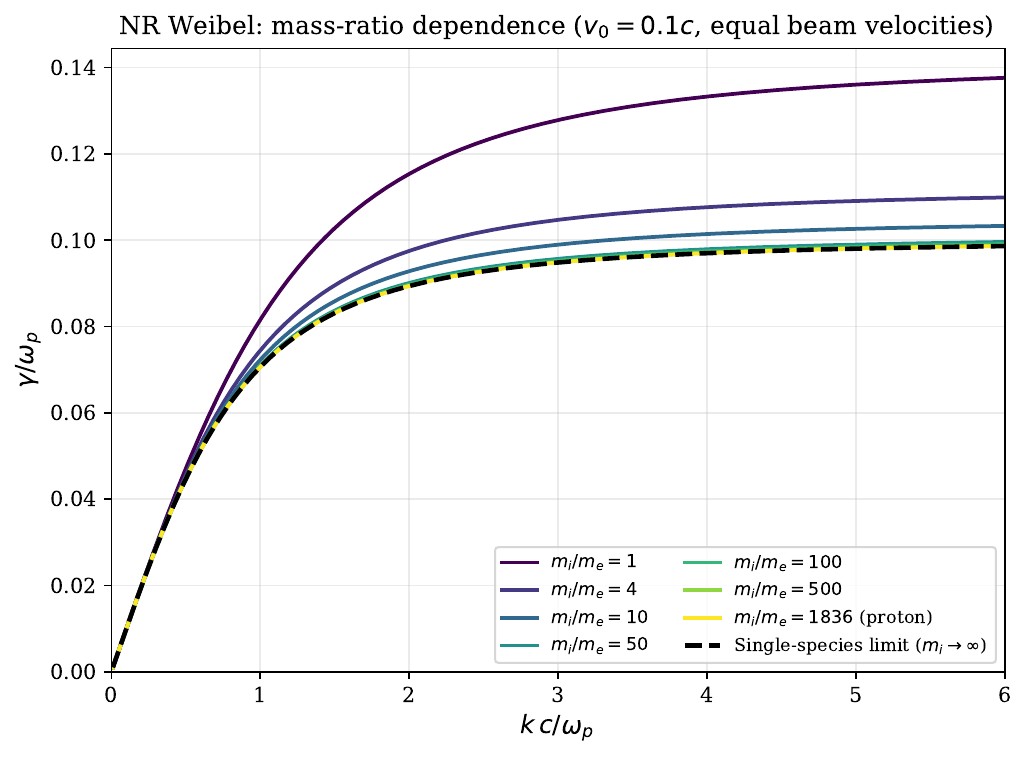}
    \caption{Non-relativistic multi-species Weibel growth rate versus
    wavenumber $k$ for mass ratios $m_i/m_e = 1$--$1836$ at $v_0 = 0.1c$,
    with equal beam velocities for both species.
    The single-species limit (black dashed) is recovered for
    $m_i/m_e \gtrsim 500$.
    The electron--positron case ($m_i/m_e = 1$) gives the highest growth
    rates because both species drive the instability with equal strength.}
    \label{fig:massratio}
\end{figure}

Figure~\ref{fig:massratio} shows the growth rate in the $(k, m_i/m_e)$ plane
at $v_0 = 0.1c$.

For electron--positron plasmas ($m_e/m_i = 1$), both species contribute
equally, giving $\gamma_{\rm max} \approx \sqrt{2}\,\omega_{pe}\,(v_0/c)$.
For intermediate mass ratios ($1/100 \lesssim m_e/m_i \lesssim 1/500$),
multi-species corrections are 1--10 per cent; reduced-mass PIC simulations
in this range should use Eq.~(\ref{NR_multi_disp}).
For realistic electron--proton ratios ($m_e/m_i \lesssim 1/1836$), the
single-species approximation is accurate to within 0.1 per cent, as
confirmed by the convergence of the curves in Fig.~\ref{fig:massratio}.

\begin{figure}[ht]
    \centering
    \includegraphics[width=\linewidth]{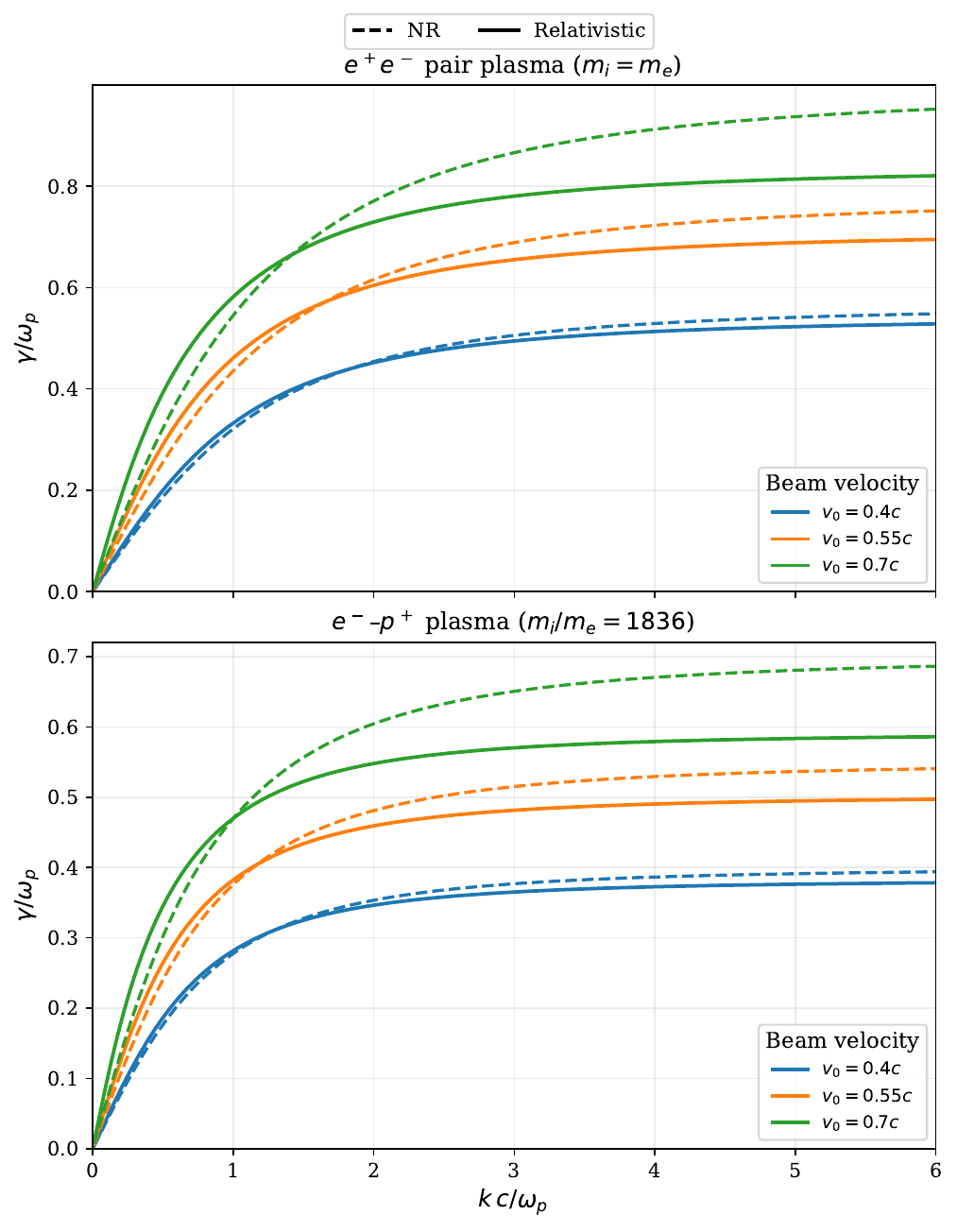}
    \caption{Growth rate $\gamma(k)$ for multi-species Weibel instability at
    beam velocities $v_0 \in \{0.4, 0.55, 0.7\}c$.
    Top panel: electron--positron pair plasma ($m_i = m_e$); both species
    contribute equally, giving higher growth rates and stronger relativistic
    suppression.
    Bottom panel: electron--proton plasma ($m_i/m_e = 1836$); results are
    close to the single-species electron limit.
    Solid: relativistic; dashed: non-relativistic.
    The pair plasma (top) consistently exceeds the electron--proton case
    (bottom) by a factor of $\approx 1.4$ at each velocity.}
    \label{fig:multispecies}
\end{figure}

Figure~\ref{fig:multispecies} compares the NR and relativistic growth rates
for electron--positron ($m_i = m_e$) and electron--proton
($m_i/m_e = 1836$) plasmas at three intermediate velocities, where
relativistic corrections are significant.

The dependence on $\omega_{pe}/\omega_{pi}$ and $v_{0e}/v_{0i}$ follows
directly from Eq.~(\ref{NR_multi_disp}): growth rate peaks near
$\omega_{pe}/\omega_{pi} = 1$ and scales quadratically with $v_{0e}$,
with the qualitative behaviour preserved in the relativistic case.

\section{\label{discussion}Discussion: Regime Boundaries and Limits of the
Theory}

\subsection{Practical Regime-Selection Criteria}

The regime boundaries follow directly from Table~\ref{tab:regimes} and
Figs.~\ref{fig:gammadiff}--\ref{fig:massratio}.

\textit{Use Eq.~(\ref{NR_single_disp})} when $v_0/c \lesssim 0.2$ and
$m_e/m_i \lesssim 1/500$.
This covers the tabletop laser-plasma shock experiment of Bai et al.\
\cite{bai2025observation, zhao2024laboratory} and young SNR environments;
growth rate errors relative to the relativistic formula are below about
4 per cent.

\textit{Use Eq.~(\ref{NR_multi_disp})} when $v_0/c \lesssim 0.2$ and
$m_e/m_i \gtrsim 1/100$. This applies to electron--positron simulations and
pair plasmas.

\textit{Use Eq.~(\ref{rel_single_disp})} when $v_0/c \gtrsim 0.2$ and
$m_e/m_i \lesssim 1/500$.
This describes ultra-relativistic electron beams and the electron-driven
Weibel precursor in GRB-scale shocks
\cite{medvedev1999generation, silva2003interpenetrating}.

\textit{Use Eq.~(\ref{rel_multi_disp})} when $v_0/c \gtrsim 0.2$ and
$m_e/m_i \gtrsim 1/100$.
The main context is electron--positron jets in GRBs and blazars.

\subsection{\label{sec:competing}Competing Instabilities and Range of
Validity}

A two-beam plasma supports multiple unstable modes.
The Bret \cite{bret2009weibel} analysis of the full three-dimensional
unstable spectrum for cold symmetric counterstreaming beams found the
following hierarchy.

In the sub-relativistic regime ($v_0 \lesssim 0.2c$) with equal beam and
background densities, the purely transverse filamentation (Weibel) mode
has the highest growth rate among electromagnetic modes.
The longitudinal two-stream instability grows at $k \parallel v_0$ but only
at scales $k \gg \omega_p/c$, well separated from the Weibel peak at
$k \sim \omega_p/c$. The Bell mode requires a background magnetic field and
does not apply here.
In this regime our analysis is self-consistent: the transverse mode we
analyse is the dominant electromagnetic instability.

In the mildly relativistic regime ($0.2c \lesssim v_0 \lesssim 0.5c$),
oblique modes can have growth rates comparable to the transverse Weibel
mode \cite{bret2009weibel, bret2010multidimensional}.
Our formulas for this regime give correct transverse-mode growth rates but
do not include competing oblique modes; a complete analysis requires the
full 3D dielectric tensor \cite{bret2009weibel}.

In the highly relativistic regime ($v_0 \gtrsim 0.5c$), the oblique (mixed)
mode typically dominates and can exceed the purely transverse Weibel mode by
a factor of two or more \cite{bret2009weibel, bret2010multidimensional}.
The relativistic multi-species results should therefore be understood as
giving the transverse-mode contribution only; a full GRB Weibel analysis
must include oblique modes, as done in kinetic simulations
\cite{silva2003interpenetrating, fiuza2012weibel}.

The density-asymmetric case ($n_{\rm beam} \ll n_{\rm bg}$), relevant to
shock-foot reflected populations, also modifies the dominant mode structure.
Bret et al.\ \cite{bret2010multidimensional} showed that for
$n_{\rm beam}/n_{\rm bg} \ll 1$, the two-stream instability can dominate at
the low beam velocities characteristic of the reflected fraction.
Our equal-density derivation should therefore be used with caution in this
limit.

\subsection{Limits of the Cold-Fluid Framework}

The cold-fluid dispersion relations predict $k_{\rm max}$ reliably because
$k_{\rm max} = \omega_{pi}/c$ follows from the structure of the dispersion
relation and is preserved qualitatively by kinetic corrections.
For $\gamma_{\rm max}$, three conditions must hold for quantitative accuracy.

\textit{First}, $v_{\rm th}/v_0 \ll 1$.
When violated, kinetic effects reduce $\gamma_{\rm max}$ significantly.
Kuldinow \& Hara \cite{kuldinow2025weibel} showed that the correction factor
scales approximately as
\begin{equation}
    \gamma_{\rm max}^{\rm kinetic} \approx \gamma_{\rm max}^{\rm cold}
    \left(1 - \frac{v_{\rm th}^2}{v_0^2}\right)^{3/2};
    \label{kinetic_correction}
\end{equation}
for the 2015 MMS crossing ($\beta_i \approx 9$), this correction gives
a suppression factor $\approx 0.08$ (factor-of-12 reduction), while for the
lower-$\beta$ 2017 event ($\beta_i \approx 2.4$), the suppression factor
is $\approx 0.67$.
For the highest-$\beta$ events analysed by
Kropotina et al.\ \cite{kropotina2023weibel}, where the background ion
distribution is nearly as energetic as the streaming population, this
correction suppresses $\gamma_{\rm max}$ by a factor of 3--5.

\textit{Second}, the instability must be in the linear phase.
Nonlinear saturation through magnetic trapping and filament coalescence
requires PIC or Vlasov simulation
\cite{fiuza2012weibel, fiuza2020electron, jikei2024saturation}.

\textit{Third}, background magnetic fields must be negligible; pre-existing
fields shift the instability threshold and preferred wavenumber
\cite{emelyanov2024weibel, burgess2016}.

\subsection{Observational Diagnostics and Inversion}

If the magnetic energy spectrum shows a filament or break at wavenumber
$k_F$, the ion inertial length is $d_i \approx 1/k_F$, giving
\begin{equation}
    n_i \approx \frac{Z_{\rm eff}^2 e^2}{\pi^2 m_i c^2}
    \left(\frac{c}{\lambda_F}\right)^2.
\end{equation}
If an e-folding time $\tau_{\rm meas}$ is known and $v_{\rm th}/v_0 \ll 1$
is confirmed, then
$\omega_{pi} \approx (v_0/c)/(0.7\,\tau_{\rm meas})$.
This growth rate inversion carries a kinetic uncertainty of order
$(v_{\rm th}/v_0)^2$ and should not be applied in warm-beam environments
without applying Eq.~(\ref{kinetic_correction})
\cite{kuldinow2025weibel}.
The scale inversion is more robust and is applicable to laser-plasma
radiography \cite{zhao2024laboratory} and spacecraft observations
\cite{stawarz2022turbulence, kropotina2023weibel}.

\section{\label{conclusions}Conclusions}

We have presented a self-contained cold-fluid analysis of the purely
transverse Weibel instability across four fluid regimes, with full
derivations, explicit regime-transition criteria, and multi-environment
comparison. The dispersion relations are known; the contribution here is
the regime-selection maps, the quantitative transition criteria, the
discussion of competing instabilities, and the comparison with the Bai
et al.\ tabletop experiment.

The main findings are as follows.

\textit{Relativistic suppression.}
Relativistic corrections are negligible below $v_0 \approx 0.2c$, become
progressively significant in the trans-relativistic regime, and reduce
$\gamma_{\rm max}$ by up to 40 per cent near $v_0 \approx 0.9c$, with the
largest suppression concentrated near $k \approx 2$--$3\,\omega_p/c$
(Fig.~\ref{fig:gammadiff}).
The relativistic growth rate peaks at $v_0 \approx 0.82c$ and decreases
at higher velocities (Fig.~\ref{fig:gammamax}).

\textit{Multi-species effects.}
Multi-species effects are significant only for $m_e/m_i \gtrsim 1/500$.
Electron--proton plasmas are well described by the single-species formulas.
Electron--positron plasmas show approximately 40 per cent higher
$\gamma_{\rm max}$ at the same beam velocity (Fig.~\ref{fig:multispecies}).

\textit{Competing instabilities.}
The transverse Weibel mode is the dominant electromagnetic instability for
$v_0 \lesssim 0.2c$. At higher velocities, oblique modes must be included;
the present formulas give the transverse-mode contribution only.

\textit{Bai et al.\ (2025) tabletop experiment.}
For the Al$^{+8}$ experiment of Bai et al.\ \cite{bai2025observation},
the non-relativistic single-species formula gives
$d_i = c/\omega_{pi} \approx 31.7\,\mu{\rm m}$, in agreement with the
observed filament spacing $\lambda_F \approx 31\,\mu{\rm m}$ to within
2 per cent.
This is the main quantitative result of the observational comparison.
The cold-fluid growth rate at the preplasma density gives
$\gamma(k_{\rm max}) \approx 4.7\times10^{11}\,{\rm s}^{-1}$
($\tau_e \approx 2.1\,{\rm ps}$ at preplasma density), providing an upper
bound on the shock formation time.
The saturation field bounds,
$B_{\rm sat} \approx 2.3\times10^4\,{\rm T}$
(Larmor and equipartition estimates coincide), bound the measured 5000~T
from above by a factor of $\approx 4.7$, consistent with kinetic
suppression.

\textit{MMS in-situ comparison.}
Two MMS burst-mode bow shock crossings were analysed using PySPEDAS
FGM and FPI-DIS data (Figs~\ref{fig:mmstsA}--\ref{fig:discatter}).
For Event~1 (2015-10-16, $\beta_i = 9.0$), the upstream $d_i = 50.8$~km
coincides with the spectral break in $P_{B_\perp}$ to within 2 per cent;
the high $\beta_i$ indicates that kinetic suppression significantly reduces
$\gamma_{\rm max}$ below the cold-fluid bound (see Eq.~\ref{kinetic_correction}
and Section~\ref{discussion}).
For Event~2 (2017-11-25, $\beta_i = 2.4$), $d_i = 61.8$~km and the
$B_\perp$ spectrum steepens at $kd_i \approx 1$ consistent with the
prediction; the lower $\beta_i$ implies more moderate kinetic suppression
of the growth rate.
Both events and the Bai et al.\ laser-plasma point lie within a factor of 3
of the 1:1 line on the multi-environment $d_i$ scatter plot
(Fig.~\ref{fig:discatter}), confirming $k_{\rm max} = \omega_{pi}/c$ across
21 orders of magnitude in ion density, consistent with
Kropotina et al.\ \cite{kropotina2023weibel}.

\textit{Framework limitations.}
The cold-fluid framework provides reliable scale predictions
($k_{\rm max}$) and upper bounds on $\gamma_{\rm max}$.
It is not adequate for quantitative growth rate predictions in warm-beam
environments.
Kinetic corrections \cite{kuldinow2025weibel, skoutnev2019temperature}
and nonlinear saturation
\cite{fiuza2012weibel, jikei2024saturation, kropotina2023weibel} are
needed for complete predictions.
Future extensions should include finite-temperature pressure-tensor
corrections \cite{kuldinow2025weibel} and background magnetic field
effects \cite{emelyanov2024weibel}.
Independent PIC verification of the regime-transition thresholds, in
particular the $v_0/c = 0.2$ non-relativistic boundary and the
$m_e/m_i = 1/500$ multi-species threshold, is left for future work.

\section*{Acknowledgements}

Authors VS, MKC and BS acknowledge the University Grants Commission (UGC),
India, for supporting this work through a Non-NET fellowship.
VS and RM thank James Juno and Ammar Hakim of the Princeton Plasma Physics
Laboratory and Petr Cagas of Virginia Tech for their suggestions and
discussions. RM also acknowledges the Visiting Associateship Program of the
Inter-University Centre for Astronomy and Astrophysics (IUCAA).
All computations were carried out on the `Brahmagupta' high-performance
computing facility at Sikkim University.

\section*{References}
\bibliographystyle{unsrt}
\bibliography{references}

\end{document}